\documentclass[fleqn]{article}
\usepackage{amsfonts,amssymb,cite}
\usepackage{graphicx}

\def\noi{\noindent}

\def\Title#1{\noi {{\Large\bf #1}}\\[1ex]}

\def\Aunames#1{\noi{\bf #1}}
\def\auth#1{${}^{#1}$}
\def\Addresses#1{\medskip\noi \protect
    \begin{description}\itemsep -3pt {\small\it #1} \end{description}}
\def\addr#1#2{\item[${}^{#1}$]{\it #2}}

\def\Abstract#1{\vskip 2mm \begin{center}
        \parbox{16.4cm}{\small\noi #1} \end{center}\medskip}
\def\PACS#1{\begin{center}{\small PACS: #1}\end{center}}

\def\email#1#2{\footnotetext[#1]{e-mail: #2}\addtocounter{footnote}{1}}

\def\nqq{\hspace*{-2em}}
\def\nhq{\hspace*{-0.5em}}

\def\cm{\hspace*{1cm}}
\def\inch{\hspace*{1in}}

\def\Jl#1#2{#1 {\bf #2},\ }

\def\ApJ#1 {\Jl{Astroph. J.}{#1}}
\def\CQG#1 {\Jl{Class. Quantum Grav.}{#1}}
\def\DAN#1 {\Jl{Dokl. AN SSSR}{#1}}
\def\GC#1 {\Jl{Grav. Cosmol.}{#1}}
\def\GRG#1 {\Jl{Gen. Rel. Grav.}{#1}}
\def\JETF#1 {\Jl{Zh. Eksp. Teor. Fiz.}{#1}}
\def\JETP#1 {\Jl{Sov. Phys. JETP}{#1}}
\def\JHEP#1 {\Jl{JHEP}{#1}}
\def\JMP#1 {\Jl{J. Math. Phys.}{#1}}
\def\NPB#1 {\Jl{Nucl. Phys. B}{#1}}
\def\NP#1 {\Jl{Nucl. Phys.}{#1}}
\def\PLA#1 {\Jl{Phys. Lett. A}{#1}}
\def\PLB#1 {\Jl{Phys. Lett. B}{#1}}
\def\PRD#1 {\Jl{Phys. Rev. D}{#1}}
\def\PRL#1 {\Jl{Phys. Rev. Lett.}{#1}}

\def\al{&\nhq}                        
\def\lal{&&\nqq {}}                   
\def\eq{Eq.\,}                        \def\nnn{\nonumber\\ \lal }
\def\eqs{Eqs.\,}                      
\def\beq{\begin{equation}}            \def\yy{\\[5pt] {}}
\def\eeq{\end{equation}}              
\def\bear{\begin{eqnarray}}           \def\eql{\al =\al}
\def\bearr{\begin{eqnarray} \lal}     
\def\ear{\end{eqnarray}}              
\def\earn{\nonumber \end{eqnarray}}

               \def\tst{\textstyle}
   \def\fract#1#2{{\tst\frac{#1}{#2}}}
             \def\half{{\fract{1}{2}}}

\def\e{{\,\rm e}}                     \def\sign{\mathop{\rm sign}\nolimits}
\def\d{\partial}                      \def\diag{\mathop{\rm diag}\nolimits}
     
     \def\const{{\rm const}}

\newcommand{\toas}{\mathop {\ \longrightarrow\ }\limits }

\newcommand{\vars}[1]{\left\{\begin{array}{ll}#1\end{array}\right.}

\def\({\left(}
\def\){\right)}

\newcommand{\Ref}[1]{(\ref{#1})}

\def\MN{^{\mu\nu}}
\def\mN{_\mu^\nu}

\def\R{{\mathbb R}}

\def\wh{wormhole}
\def\whs{wormholes}

\def\ssph{static, spherically symmetric}
\def\asflat{asymptotically flat}

\begin{document}
\twocolumn[

\Title{Trapped ghosts: a new class of wormholes}

\Aunames {Kirill A. Bronnikov\auth{a,1}, and Sergey V. Sushkov\auth{b,2}
      }

\Addresses{
\addr a {Center of Gravitation and Fundamental Metrology,
     VNIIMS, Ozyornaya St. 46, Moscow 117361, Russia; \\
     Institute of Gravitation and Cosmology,
         PFUR, Miklukho-Maklaya St. 6, Moscow 117198, Russia  }
\addr b
{Department of General Relativity and Gravitation, Kazan State University,\\
    Kremlyovskaya St. 18, Kazan 420008, Russia;\\
 Department of Mathematics, Tatar State University of Humanities and
    Education,\\ Tatarstan St. 2, Kazan 420021, Russia}
      }

\Abstract
  {We construct examples of \ssph\ \wh\ solutions in general relativity with
   a minimally coupled scalar field $\phi$ whose kinetic energy is negative
   in a restricted region of space near the throat (of arbitrary size) and
   positive far from it. Thus in such configurations a ``ghost'' is trapped
   in the strong-field region, which may in principle explain why no ghosts
   are observed under usual conditions. Some properties of general \wh\
   models with the $\phi$ field are revealed: it is shown that (i)
   trapped-ghost \whs\ are only possible with nonzero potentials $V(\phi)$;
   (ii) in twice \asflat\ \whs, a nontrivial potential $V(\phi)$ has an
   alternate sign, and (iii) a twice \asflat\ \wh\ which is mirror-symmetric
   with respect to its throat has necessarily a zero Schwarzschild mass at
   both asymptotics.  }


] %%%%%%%%%%%%%%%%%%%%%%%%%%%%%%%%%%%%%%%%%%%%%%%%%%%%%%
\email 1 {kb20@yandex.ru}
\email 2 {sergey\_sushkov@mail.ru}

\section{Introduction}

  It is well known that the existence of traversable Lo\-rentz\-ian \whs\ as
  solutions to the equations of general relativity requires ``exotic
  matter'', i.e., matter that violates the null energy condition
  \cite{thorne, hoh-vis}. In particular, for \ssph\ configurations with a
  minimally coupled scalar field as a source, \wh\ solutions are only
  possible if the scalar field is phantom, i.e., has a wrong sign of kinetic
  energy \cite{br73, vac1,SusZha}. In alternative theories of gravity, such as
  scalar-tensor, multidimensional and curvature-nonlinear theories, \whs\
  also turn out to be possible only if some of the degrees of freedom are of
  phantom nature \cite{br73, bstar07} (see also the review \cite{lobo} and
  references therein).

  Meanwhile, macroscopic phantom matter has not yet been observed, which puts
  to doubt the very possibility of obtaining realistic \whs\ even in a remote
  future and even by a highly advanced civilization.

  In this paper, we would like to discuss an interesting opportunity of
  obtaining \wh\ configurations in general relativity with a kind of matter
  which possesses phantom properties only in a restricted region of space,
  somewhere close to the throat, whereas far away from it all standard
  energy conditions are observed. As an example of such matter, we consider
  \ssph\ configurations of a minimally coupled scalar field with the
  Lagrangian\footnote
      {We choose the metric signature ($-,+,+,+$), the units $c = \hbar =
       8\pi G=1$, and the sign of $T\mN$ such that $T^0_0$ is the energy
       density.}
\beq                                                         \label{L_s}
      L_s = -\half h(\phi) g\MN \d_\mu \phi\d_\nu \phi - V(\phi),
\eeq
  where $h(\phi)$ and $V(\phi)$ are arbitrary functions. If $h(\phi)$ has a
  variable sign, it cannot be absorbed by re-definition of $\phi$ in its
  whole range. A case of interest is that $h >0$ (that is, the scalar field
  is canonical, with positive kinetic energy) in a weak field region and
  $h < 0$ (the scalar field is of phantom, or ghost nature) in some
  restricted region where a \wh\ throat can be expected. In this sense it can
  be said that the ghost is trapped. A possible transition between $h > 0$
  and $h < 0$ in cosmology was considered in \cite {rubin}.

  The paper is organized as follows. In Section 2 we present the basic
  equations and show why the above strategy cannot be realized for massless
  fields ($V(\phi)\equiv 0$). In Section 3 we describe some general
  properties of the system and obtain explicit examples of ``trapped-ghost''
  \wh\ solutions using the inverse-problem method, and Section 4 is a
  conclusion.

\section{Scalar fields with a variable kinetic term}

  The general static, spherically symmetric metric can be written as
\beq                                                           \label{ds}
        ds^2 = -\e^{2\gamma(u)}dt^2 + \e^{2\alpha(u)}du^2
                                  + \e^{2\beta(u)} d\Omega^2.
\eeq
  where $u$ is an arbitrary radial coordinate and $d\Omega^2 = (d\theta^2 +
  \sin^2\theta d\varphi^2)$ is the linear element on a unit sphere.
  A scalar field $\phi(u)$ with the Lagrangian (\ref{L_s}) in a space-time
  with the metric (\ref{ds}) has the stress-energy tensor (SET)
\bearr
     T\mN = \half h(u) \e^{-2\alpha} \phi'(u)^2 \diag (1,\ -1,\ 1,\ 1)
\nnn  \inch
        + \delta\mN V(u).                                       \label{SET}
\ear
  (the prime denotes $d/du$). The kinetic energy density is positive if
  $h(\phi) >0$ and negative if $h(\phi) < 0$, so \wh\ solutions sought for
  must be obtained with $h >0$ at large values of the spherical radius $r(u)
  = \e^{\beta}$ and $h < 0$ at smaller radii $r$. One can show, however, that
  this goal cannot be achieved for a massless field ($V(\phi) \equiv 0$).

  Indeed, in the massless case, the SET (\ref{SET}) has the same structure as
  for a usual massless scalar field with $h = \pm 1$. Therefore the metric
  has the same form as in this simple case and should be reduced to the
  Fisher metric \cite{Fis} if $h > 0$ and to the corresponding solution for a
  phantom scalar, first found by Bergmann and Leipnik \cite{BerLei} (it is
  sometimes called ``anti-Fisher'') in case $h(\phi) < 0$. Let us reproduce
  this solution for our scalar (\ref{L_s}) in the simplest joint form,
  following \cite{br73}.

  Two combinations of the Einstein equations
\beq                                                             \label{EE}
    R\mN - \half \delta\mN R = - T\mN
\eeq
  for the metric (\ref{ds}) and the SET (\ref{SET}) with $V\equiv 0$ read
  $R^0_0 =0$ and $R^0_0 + R^2_2 =0$.  Choosing the harmonic radial coordinate
  $u$, such that $\alpha(u) = 2\beta(u) + \gamma(u)$, we easily solve these
  equations. Indeed, the first of them reads simply $\gamma'' =0$, while the
  second one is written as $\beta'' + \gamma'' = \e^{2(\beta+\gamma)}$.
  Solving them, we have
\bearr
     \gamma = - mu,                                              \label{s}
\nnn
     \e^{-\beta-\gamma} =  s(k,u) := \vars     {
                    k^{-1}\sinh ku,  \ & k > 0, \\
                                 u,  \ & k = 0, \\
                    k^{-1}\sin ku,   \ & k < 0.     }
\ear
  where $k$ and $m$ are integration constants; two more integration constants
  have been suppressed by choosing the zero point of $u$ and the scale along
  the time axis. As a result, the metric has the form \cite{br73}\footnote
    {For a detailed description of the properties of Fisher and anti-Fisher
     solutions see \cite{cold08, SusZha, sigma} and references therein.}
\beq                                                           \label{ds1}
     ds^2 = -\e^{-2mu} dt^2 + \frac{\e^{2mu}}{s^2(k,u)}
                    \biggr[\frac{du^2}{s^2(k,u)} + d\Omega^2\biggl]
\eeq
  (note that spatial infinity here corresponds to $u=0$ and $m$ has the
  meaning of the Schwarzschild mass). Moreover, with these metric functions,
  the ${1\choose 1}$ component of the Einstein equations (\ref{EE}) leads to
\beq                                                           \label{int}
     k^2 \sign k = m^2 + \half h(\phi) \phi'^2.
\eeq
  It means that $h(\phi) \phi'^2 = \const$, that is, $h(\phi)$ cannot change
  its sign within a particular solution which is characterized by certain
  values of the constants $m$ and $k$. The situation remains the same if,
  instead of a single scalar field, there is a nonlinear sigma model with
  multiple scalar fields $\phi^a$ and the Lagrangian
\beq                                                           \label{L-sigma}
      L_\sigma = -\half h_{ab} g\MN \d_\mu \phi^a\d_\nu \phi^b,
\eeq
  where $h_{ab}$ are functions of $\phi^a$: the metric then has the same form
  (\ref{ds1}), and a relation similar to (\ref{int}) reads \cite{sigma}
\[
      k^2 \sign k = m^2 + \half h_{ab} \phi^a{}' \phi^b{}'.
\]
  Therefore the quantity that determines the canonical or phantom nature of
  the scalars, $h_{ab} \phi^a{}' \phi^b{}'$, is constant. If the matrix
  $h_{ab}$ is not positive- or negative-definite, some solutions due to
  (\ref{L-sigma}) can be \wh\ while others correspond to a canonical scalar
  and have a Fisher central singularity \cite{sigma}, but there are no
  solutions of trapped-ghost character.

  Returning to our system with the Lagrangian (\ref{L_s}), we can assert that
  trapped-ghost \whs\ can only exist with a nonzero potential $V(\phi)$.
  To find such configurations, it is helpful to use the so-called quasiglobal
  gauge $\alpha + \gamma=0$, so that the metric (\ref{ds}) takes the form
\beq
    ds^2 = - A(u) dt^2 + \frac{du^2}{A(u)} + r^2(u)d\Omega^2,    \label{ds2}
\eeq
  where $A(u)$ is called the redshift function and $r(u)$ the area function.
  Then the Einstein-scalar equations can be written as
\bear
     (A r^2 h\phi')' - \half Ar^2 h'\phi' \eql r^2 dV/d\phi,  \label{phi}
\yy
              (A'r^2)' \eql - 2r^2 V;                         \label{00}
\yy
              2 r''/r \eql - h(\phi){\phi'}^2 ;               \label{01}
\yy
         A (r^2)'' - r^2 A'' \eql 2,                          \label{02}
\yy                                                   \label{11}
      -1 + A' rr' + Ar'^2 \eql r^2 (\half h A \phi'^2 -V),
\ear
  where the prime again denotes $d/du$. \eq (\ref{phi}) follows from
  (\ref{00})--(\ref{02}), which, given the potential $V(\phi)$ and the
  kinetic function $h(\phi)$, form a determined set of equations for the
  unknowns $r(u)$, $A(u)$, $\phi(u)$. \eq (\ref{11}) (the ${1\choose 1}$
  component of the Einstein equations), free from second-order derivatives,
  is a first integral of (\ref{phi})--(\ref{02}) and can be obtained from
  (\ref{00})--(\ref{02}) by excluding second-order derivatives. Moreover,
  \eq(\ref{02}) can be integrated giving
\bear
            B'(u) \equiv (A/r^2)' = 2(3m - u)/r^4,          \label{B'}
\ear
  where $B(u) \equiv A/r^2$ and $m$ is an integration constant equal to the
  Schwarzschild mass if the metric (\ref{ds}) is \asflat\ as $u\to \infty$
  ($r \approx u$, $A = 1 - 2m/u + o(1/u)$).
  If there is a flat asymptotic as $u\to -\infty$, the Schwarzschild
  mass there is equal to $-m$ ($r \approx |u|$, $A = 1 + 2m/|u| + o(1/u)$.

  Thus in any \wh\ solution with two flat asymptotics we inevitably have
  masses of opposite signs, just as is the case in the well-known special
  solution --- the anti-Fisher \wh\ \cite{ellis, cold08, SusZha} whose metric
  in the gauge (\ref{ds2}) reads
\[
     ds^2 = -\e^{-2mz} dt^2 + \e^{2mz} [du^2 + (k^2 + u^2) d\Omega^2],
\]
  with $k < 0$ and $z = |k|^{-1} \cot^{-1} (u/|k|)$ [the constants $m$ and $k$
  have the same meaning as in (\ref{s})--(\ref{int})].

  It is also clear that $m = 0$ in all symmetric solutions to \eqs
  (\ref{phi})--(\ref{11}), such that $r(u)$ and $A(u)$ are even functions.
  Indeed, in this case $B'(u)$ is odd, hence $m=0$ in (\ref{B'}).

\section{Models with a trapped ghost}

  If one specifies the functions $V(\phi)$ and $h(\phi)$ in the Lagrangian
  (\ref{L_s}), it is, in general, hard to solve the above equations.
  Alternatively, to find examples of solutions possessing some particular
  properties, one may employ the inverse problem method, choosing some of the
  functions $r(u)$, $A(u)$ or $\phi(u)$ and then reconstructing the form of
  $V(\phi)$ and/or $h(\phi)$. We will do so, choosing a function $r(u)$ that
  describes a \wh\ profile. Then $A(u)$ is found from (\ref{B'}) and $V(u)$
  from (\ref{00}). The function $\phi(u)$ is found from (\ref{01}) provided
  $h(\phi)$ is known; however, using the scalar field parametrization
  freedom, we can, vice versa, choose a monotonic function $\phi(u)$ (which
  will yield an unambiguous function $V(\phi)$) and find $h(u)$ from \eq
  (\ref{01}).

  Let us discuss what kind of function $r(u)$ is required for our purpose.

\begin{enumerate}
\item
     The \wh\ throat ($u=0$ without loss of generality) is a minimum of
     $r(u)$, that is,
\beq
        r(0) = a, \qquad r'(0) =0, \qquad r''(0) > 0
\eeq
     with $a=\const >0$ (these requirements are sometimes called the
     flare-out conditions). So $r(u)$ must have such a minimum.
\item
     In a trapped-ghost wormhole, by definition, the kinetic coupling
     function $h(\phi)$ is negative near the throat and positive far from it.
     According to (\ref{01}), this means that $r''$ is positive at small
     $|u|$ and negative at sufficiently large $|u|$.
\item
     If the wormhole is asymptotically flat at large $|u|$, we should have
\[
        r(u) \approx |u|\qquad {\rm as}\qquad  u\to \pm \infty.
\]
\end{enumerate}

\begin{figure}[h]
\centerline{\includegraphics[width=8cm]{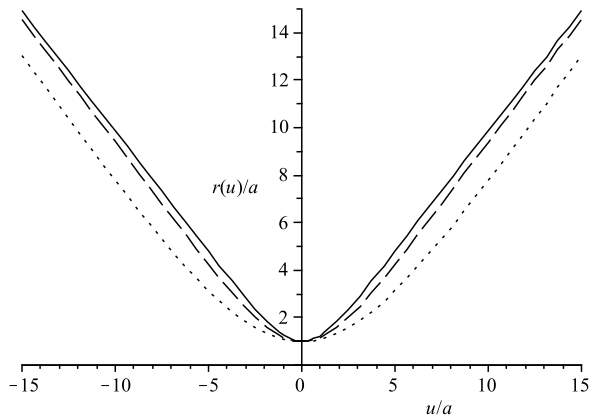}}
\caption{Plots of $r(u)/a$ given by \eq \Ref{r} with $\lambda=3; 5; 10$
    (solid, dashed, and dotted lines, respectively).
\label{figr}}
\end{figure}

  A simple example of the function $r(u)$ satisfying the requirements
  1--3 is (see Fig. \ref{figr}):
\beq\label{r}
         r(u) = a \frac{(x^2+\lambda)} {\sqrt{x^2+\lambda^2}},\cm
       \lambda = \const > 2.
\eeq
  where $x = u/a$, and $a$ is the (arbitrary) throat radius.

\begin{figure}[h]
\centerline{\includegraphics[width=8cm]{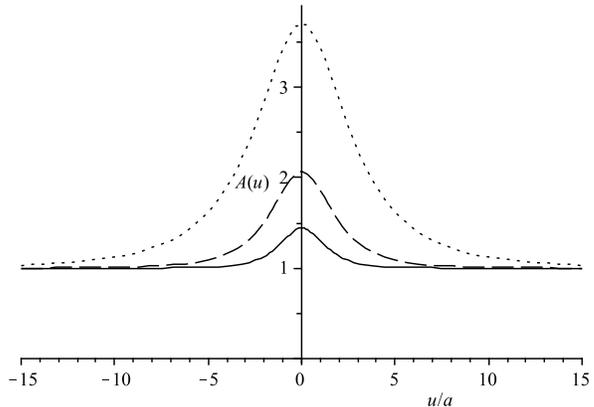}}
\caption{Plots of $A(u)$ given by \eq \Ref{A} with $\lambda=3; 5; 10$ (solid,
     dashed, and dotted lines, respectively). \label{figA}}
\end{figure}

  Now we can integrate \eq\Ref{B'}. Assuming $m=0$, we find (see
  Fig.\,\ref{figA})
\bear\label{A}
    A(u) = \frac{3x^4+3\lambda(\lambda+1)x^2
        +\lambda^2(\lambda^2+\lambda+1)}{3(x^2+\lambda)(x^2+\lambda^2)}.
\ear

  Substituting the expressions \Ref{r} and \Ref{A} into \Ref{00}, we
  obtain the potential $V$ as a function of $u$ or $x = u/a$:
\bearr\label{V}
    V(u) = \frac{\lambda^2(\lambda-1)^2
            [-6x^4+\lambda(\lambda-5)x^2+\lambda^3(\lambda+1)]}
                    {3a^2 (x^2+\lambda)^2 (x^2+\lambda^2)^3}.
\nnn
\ear
  One can notice that $V(u) < 0$ at large $|u|$. The negative sign of the
  potential in a certain range of $u$ is not a shortcoming of this
  particular model but a direct consequence of the field equations. Indeed,
  as follows from (\ref{B'}), we have  $A'r^2 \toas_{u\to\pm\infty} 2m$
  at both flat aymptotics. Consequently, due to (\ref{00}),
\[
    \int_{-\infty}^{+\infty} r^2 V(u) du = 0,
\]
  so that if $V(u)\not \equiv 0$, it has an alternate sign.

\begin{figure}[ht]
\centerline{\includegraphics[width=8cm]{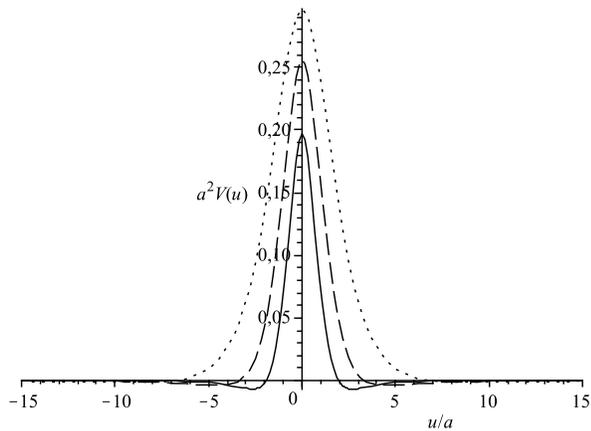}} \caption{Plots of
	$V(\phi)$ given by \eq \Ref{V} with $\lambda=3; 5; 10$
	(solid, dashed, and dotted lines, respectively).
\label{figV}}
\end{figure}

  To construct $V$ as an unambiguous function of $\phi$ and to find
  $h(\phi)$, it makes sense to choose a monotonic function $\phi(u)$.
  It is convenient to assume
\beq\label{phi_2}
    \phi(u) = \frac{2\phi_0}{\pi}\arctan\frac{x}{\lambda},
    \qquad
    \phi_0 = \frac{\pi a}{2}\sqrt{\frac{2(\lambda-2)}{\lambda}},
\eeq
  and $\phi$ has a finite range: $\phi \in (- \phi_0, \phi_0)$, which is
  common to kink configurations. Thus we have $x = u/a =
  \lambda\tan(\pi\phi/2\phi_0) $, whose substitution into \eq \Ref{V} gives
  an expression for $V(\phi)$ defined in this finite range. The function
  $V(\phi)$ can be extended to the whole real axis, $\phi \in \R$, by
  supposing $V(\phi)\equiv 0$ at $|\phi| \geq \phi_0$. Plots of the extended
  potential $V(\phi)$ are shown in Fig.\,\ref{figV}.

\begin{figure}[ht]
\centerline{\includegraphics[width=8cm]{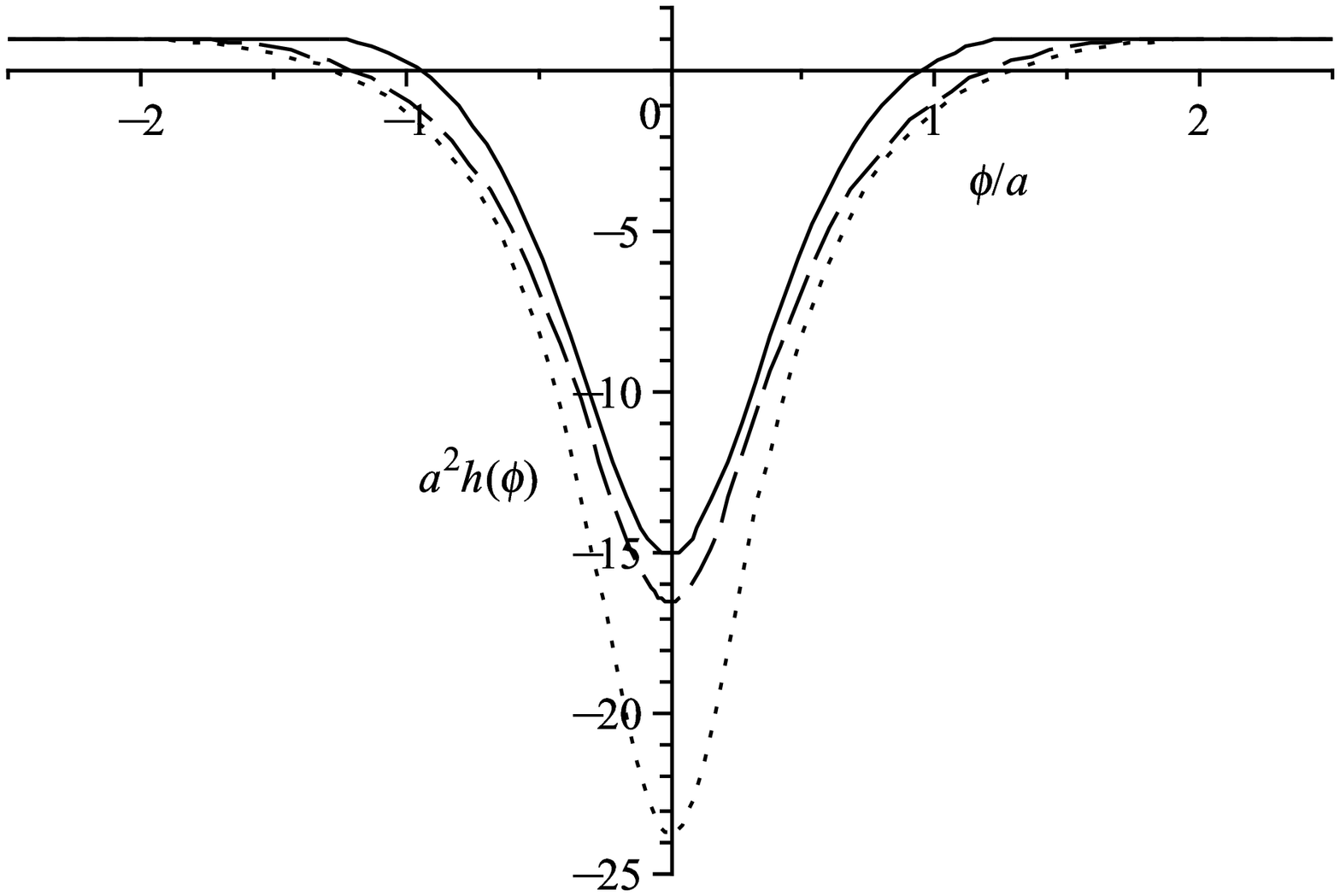}} \caption{Plots of
	$h(\phi)$ given by \eq \Ref{h2} with $\lambda=3; 5; 10$
        (solid, dashed, and dotted lines, respectively). \label{figh}}
\end{figure}

  The expression for $h(\phi)$ is found from \Ref{01} as follows:
\beq\label{h2}
    h(\phi) = \frac{(\lambda-2)x^2+\lambda^2(1-2\lambda)}
            {a^2(\lambda-2)(x^2+\lambda)},
\eeq
  where $x = \lambda\tan(\pi\phi/2\phi_0)$. The function $h(\phi)$ given
  by \eq \Ref{h2} is also defined in the interval $(-\phi_0,\phi_0)$ and
  can be extended to $\R$ by supposing $h(\phi)\equiv 1$ at
  $|\phi|\geq \phi_0$. The extended kinetic coupling function
  $h(\phi)$ is plotted in Fig.\,\ref{figh}.

  It is well known that the null energy condition (NEC) holds for a canonical
  scalar field and is violated for a phantom one. In our case, it happens for
  $h(\phi) > 0$ and $h(\phi) < 0$, respectively. Let us illustrate this using
  our solution (\ref{r}), (\ref{A}), (\ref{phi_2}) as an example. The NEC
  reads $-T_{\mu\nu}k^\mu k^\nu\ge 0$, where $k^\mu$ is an arbitrary null
  vector. Due to the Einstein equations \Ref{EE}, it can be equivalently
  written as $G_{\mu\nu}k^{\mu}k^{\nu}\ge 0$. Taking the radial null vector
  $k^\mu=(A^{-1/2},A^{1/2},0,0)$ in the metric \Ref{ds2} and denoting
  $\Xi: = G_{\mu\nu}k^{\mu}k^{\nu}$, we find
\bearr
    \Xi(u) = -2\frac{r''}{r}
    = \frac{2\lambda[x^2(\lambda-2)-\lambda^2(2\lambda-1)]}
    			{a^2(x^2+\lambda^2)(x^2+\lambda)},     \label{Xi}
\ear
  i.e., it is a multiple of $h(\phi)$ with a positive factor. The plot of
  $h(\phi)$ thus completely characterizes NEC violation taking place near 
  the throat.

\section {Conclusion}

  We have shown that a minimally coupled scalar field may change its nature
  from canonical to ghost in a smooth way without creating any space-time
  singularities. This feature, in particular, allows for construction of
  \ssph\ \wh\ models (trapped-ghost \whs) where the ghost is present in some
  restricted region around the throat (of arbitrary size) whereas in the
  weak-field region far from it the scalar has usual canonical properties.
  One can speculate that if such ghosts do exist in Nature, they are all
  confined to strong-field regions (``all genies are sitting in their
  bottles''), but just one of them, having been released, has occupied the
  whole Universe and plays the part of dark energy (if dark energy is
  really phantom, which is more or less likely but not certain).

  We have also found some general properties of \ssph\ \wh\ models in the
  Einstein-scalar field system under consideration:
\begin{description}
\item[(i)]
    trapped-ghost \whs\ are only possible with nonzero potentials $V(\phi)$;
\item[(ii)]
    in all \whs\ with two flat asymptoyics, $V(\phi)$ has an alternate sign
    (unless $V \equiv 0$);
\item[(iii)]
    in any \wh\ with two flat asymptoyics, if the Schwarzschild mass equals
    $m$ at one of them, it equals $-m$ at the other. Hence mirror symmetry
    with respect to the throat (i.e., the metric functions are even in the
    radial coordinate $u$) implies $m = 0$.
\end{description}

\subsection*{Acknowledgments}
    This work was supported in part by the Russian Foundation for
    Basic Research grants No. 08-02-91307, 08-02-00325, and 09-02-00677a.

\end{document}